\documentclass[envcountsame,runningheads]{llncs}

\usepackage[T1]{fontenc}
\usepackage{graphicx}
\usepackage{amsfonts}       
\usepackage{amsmath}		
\usepackage{cleveref}       
\usepackage{mathtools}
\usepackage[ruled,lined,linesnumbered, noend]{algorithm2e}
\usepackage[dvipsnames]{xcolor}
\usepackage{subcaption}

\usepackage{longtable}
\usepackage{multirow}
\usepackage{array}
\usepackage{nowidow}
\usepackage[ruled,lined,linesnumbered, noend]{algorithm2e}
\DontPrintSemicolon{}
\newcommand{\state}[1]{$#1$ \;}
\newcommand{\assign}[2]{\state{#1 \es{} \leftarrow{} \es{} #2}}
\newcommand{\es}{\textbf{ }}

\usepackage{enumitem}

\newcommand{\tabmargin}{0.25em}
\newcolumntype{P}[1]{>{\hspace{\tabmargin}}p{#1}<{\hspace{\tabmargin}}}
\newcolumntype{L}{>{\hspace{\tabmargin}}l<{\hspace{\tabmargin}}}
\newcolumntype{C}{>{\hspace{\tabmargin}}c<{\hspace{\tabmargin}}}
\newcolumntype{R}{>{\hspace{\tabmargin}}r<{\hspace{\tabmargin}}}

\usepackage{pscproc2}

\DeclareMathOperator{\rank}{rank}
\DeclareMathOperator{\select}{select}
\DeclareMathOperator{\degree}{degree}

\DeclareMathOperator{\node}{node}
\DeclareMathOperator{\edge}{edge}
\DeclareMathOperator{\contains}{contains}
\DeclareMathOperator{\argexists}{exists}

\DeclareMathOperator{\decompress}{decompress}
\DeclareMathOperator{\deleteEdge}{deleteEdge}
\DeclareMathOperator{\removeNode}{removeNode}
\DeclareMathOperator{\addEdge}{addEdge}
\DeclareMathOperator{\insertNode}{insertNode}
\DeclareMathOperator{\addNode}{addNode}
\DeclareMathOperator{\deleteNode}{deleteNode}

\newcommand{\bigO}{\mathcal{O}}
\newcommand{\sidenode}[1]{}

\newcommand{\href}[2]{#2}

\begin{document}

\title{Updating Permutation-Compressed Hypergraphs}
\author{Enno Adler \and
Stefan Böttcher \and
Rita Hartel}
\authorrunning{Adler et al.}
\titlerunning{Updating HyperCSA}
\institute{Paderborn University,\\ 
Warburger Straße 100,
Paderborn, Germany\\
\email{\{enno.adler, stefan.boettcher, rita.hartel\}@uni-paderborn.de}
}
\maketitle

\begin{abstract}
    Hypergraphs model complex, non-binary relationships such as co-author\-ships, social group memberships, and recommendation systems. Like traditional graphs, hypergraphs can grow large, posing challenges for storage, transmission, and query performance. Moreover, the data can change as new papers are published, new social groups form, existing social groups accept or remove members, and new products are published. In this paper, we extend the hypergraph compression method HyperCSA to perform updates of the compressed hypergraph. Dynamic HyperCSA computes the exact compressed hypergraph representation that the steps \emph{decompress-update-compress} of HyperCSA produce, but is orders of magnitude faster. Thereby, dynamic HyperCSA achieves the same compression ratios of $26\%$ to $79\%$ as HyperCSA on the modified, real-world hypergraphs and the query performance of being $6$ to $40$ times faster on real-world hypergraphs than other approaches.
\end{abstract}



\section[Introduction]{Introduction}

Hypergraphs are powerful tools for modeling complex, non-binary relationships in data. Unlike traditional graphs, where edges connect only two nodes, hypergraphs allow edges to connect any number of nodes, enabling a more accurate representation of real-world systems. This modeling flexibility has made hypergraphs useful in a range of domains, including social networks~\cite{tan2014}, music~\cite{tan2011} and news~\cite{li2013} recommendation systems, academic authorship networks, and tagged data repositories from online platforms~\cite{ghoshal2009}. To store and query large datasets, Adler et al.~\cite{adler2026} recently introduced HyperCSA, a permutation-based hypergraph compression scheme based on the compressed suffix array by Sadakane~\cite{sadakane2003}. However, HyperCSA cannot modify the underlying hypergraph yet. In this paper, we show how to alter the compressed representation of HyperCSA to reflect changes in the hypergraph data.

Herein, our main contributions are as follows: 
\begin{itemize}
    \item Algorithms\footnote{An implementation of all algorithms proposed in this paper is available at \url{https://github.com/adlerenno/hypercsa}.} to add and remove nodes or edges from a hypergraph and algorithms to add and remove a single node from an edge in a hypergraph compressed by HyperCSA.
    \item Theoretical worst-case runtimes for all mentioned algorithms.
    \item A practical evaluation showing that the proposed algorithms significantly outperform the naive decompress-update-compress approach of updating a hpergraph compressed with HyperCSA.
\end{itemize}

\section[Preliminaries]{Preliminaries}\label{section:preliminaries}

\begin{figure*}[!hbp]
    \centering
    \includegraphics[scale=0.8]{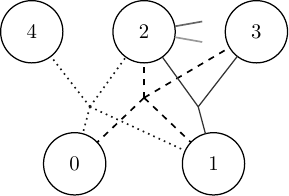}
    \caption{The example hypergraph $H$ with nodes $V_H = \{0, 1, 2, 3, 4\}$ and hyperedges $E_H = \left\{\{0, 1, 2, 3\},\es\{1, 2, 3\},\es\{2\},\es\{0, 1, 2, 4\},\es\{2\}\right\}$.}\label{figure:example_hypergraph}
\end{figure*}

A \textit{hypergraph} is a pair $G = (V_G, E_G)$ with nodes $V_G = \{0, 1, \ldots, p\} \subset \mathbb{N}_0$ and edges $E_G = \{e_0, \dots, e_{m-1}\}$. For every hyperedge $e_i = \{v_0, \ldots, v_{n-1}\} \subseteq V_G$, $e_i \not= \emptyset$ and $\rank(e_i) = |e_i| = n$. We call all hyperedges in $E_G$ edges regardless of their rank. 
HyperCSA can also handle hypergraphs that contain the same edge multiple times. In Figure~\ref{figure:example_hypergraph}, we visualize our ongoing example.

We define $\degree(v) = |\{e \in E_G : v \in e \}|$, $R_G = \max\{\rank(e): e\in E_G\}$, $\Delta_G = \max\{\degree(v):v\in V_G\}$, and $M_G = \sum_{e \in E_G} \rank(e) = \sum_{v \in V} \degree(v)$. For the example hypergraph $H$, we have $\degree(2) = 5$, $R_H = 4$, $\Delta_H = 5$, and $M_H = 13$.

A string $T$ of length $|T|$ over an alphabet $\Sigma$ with total order $<$ is a sequence of characters $T[i] \in \Sigma$ for $0 \leq i < |T|$, so $T = T[0] \cdot T[1] \cdots T[|T|-1]$. 
We write $\epsilon$ for the empty string, $T[i, j]$ for the substring $T[i] \cdot T[i+1] \cdots T[j]$ of $T$, $T[i,j] = \epsilon$ if $i > j$, and $T[i..] = T[i, |T|-1]$ for the suffix starting at position $i$. We call a position $i$ an S-type, if $T[i] < T[i+1]$, and L-type otherwise, also if $i=|T|-1$.

We define $\rank_T(x, c) = |\{i < x : T[i] = c\}|$ as the number of times $c$ occurs in the first $x$ characters of $T$ (excluding the current position). 
Next, let $\select_T(i, c) = \min\{x < |T| : \rank_T(x, c) = i\}-1$ be the position of the $i$-th $c$ in $T$. 

The suffix array $SA$~\cite{manber90} of a string $T$ is a permutation of $\{0, \ldots, |T|-1\}$ such that the $i$-th smallest suffix of $T$ is $T[SA[i]..]$ with $i \in \{0, \ldots, |T|-1\}$. As permutations are invertible, we denote the inverse of the suffix array by $SA^{-1}$. We use the term \textit{T-position} for a position in $T$ and \textit{SA-position} for a position in $SA$. The compressed suffix array $CSA$ by Sadakane~\cite{sadakane2003} of a string $T$ consists of a bitvector $D$ of length $|T|$ and a permutation $\Psi_{CSA}$ of $\{0, \ldots, |T|-1\}$ (and other here omitted data structures). The bitvector $D$ is a succinct representation for $T$ using the suffix array $SA$: $D[i] = 1$ marks the first suffix in the sorted suffixes of $T$ that start with a new character, so $D[i] = 1$ if and only if $T[SA[i]] > T[SA[i-1]]$ or $i = 0$. We assume that all symbols of $\Sigma$ have at least one occurrence in $T$. By this assumption, we get $T[SA[i]] = \rank_D(i, 1)-1$. $\Psi_{CSA}[i]$ stores the SA-position $j$ that points to the next T-position $SA[i]+1$ after the T-position $SA[i]$ at SA-position $i$. In one term, $\Psi_{CSA}[i] = SA^{-1}[(SA[i] + 1) \mod |T|]$. $\Psi_{CSA}$ is a cyclic permutation and by computing the ranks on $D$ for the SA-positions $i$ in order of the cycle of $\Psi_{CSA}$, we retrieve a rotation of $T$. 

\begin{figure*}
    \centering
    \includegraphics[width=1\linewidth]{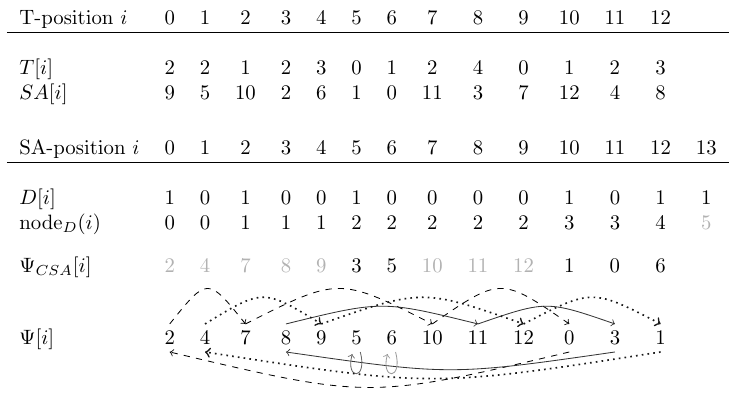}
    \caption[HyperCSA data structure of hypergraph $H$.]{HyperCSA data structure for the hypergraph $H$. The node at index $i$ is $\node_D(i) = \rank_D(i+1, 1)-1$. The black numbers in $\Psi_{CSA}$ are the numbers adjusted by the third step of HyperCSA's construction. The style of every edge in $\Psi$ is the same as in Figure~\ref{figure:example_hypergraph}. The arrows of SA-positions with $\Psi[i] \leq i$ are drawn below the numbers, all others are drawn above.}\label{figure:hypercsa:construction:psi_jumps}
\end{figure*}

The HyperCSA approach by Adler et al.~\cite{adler2026} is inspired by RDFCSA by Brisaboa et al.~\cite{brisaboa2015}, which is a self-index for the triples representing an RDF graph. As HyperCSA is a generalization of RDFCSA, all presented algorithms can be adapted for RDFCSA. HyperCSA's construction consists of three steps: First, we sort the nodes in an edge in ascending order and then, we sort the edges in lexicographically descending order. Thereafter, we concatenate all edges into one string $T$. For our example, we get $T = 2 | 2 | 1 2 3 | 0 1 2 4 | 0 1 2 3$, where the vertical bars are inserted as visual aid to identify the edges. Second, we compute the compressed suffix array of $T$ and append a $1$ to $D$ for easier querying. We show $D$ and $\Psi_{CSA}$ of the compressed suffix array for our example $T$ in Figure~\ref{figure:hypercsa:construction:psi_jumps}. Third, we modify $\Psi_{CSA}$ such that each edge gets its own cycle in the permutation $\Psi$. We depict these cycles in Figure~\ref{figure:hypercsa:construction:psi_jumps} and provide the algorithm in Agorithm~\ref{algorithm:hypercsa:construction:adjust_psi}. The adjustment from $\Psi_{CSA}$ to $\Psi$ only requires one change $\Psi_{CSA}[j] = i$ for every hyperedge $e$, where $j$ is the single $L$-type position of $e$ and $i$ is the smallest position that belongs to $e$ in $\Psi_{CSA}$.

\begin{algorithm}
    \caption{HyperCSA.$adjust\Psi$}\label{algorithm:hypercsa:construction:adjust_psi}
    \KwIn{$\Psi_{CSA}$ (in the following renamed to $\Psi$)}
    \assign{current\_position}{\Psi[0]}
    \assign{next}{\Psi[current\_position]}
    \assign{last\_first\_node\_position}{0}
    \While{$current\_position$ is not $0$}{
        \If{$next < current\_position$}{
            \assign{\Psi[current\_position]}{last\_first\_node\_position}
            \assign{last\_first\_node\_position}{next}
        }
        \assign{current\_position}{next}
        \assign{next}{\Psi[next]}
    }
\end{algorithm}

We define $\node(i) = \rank_D(i + 1, 1) - 1$ and $\edge(i) = \{\node_D(\Psi^k[i]) : k = 0, 1, \dots\}$. We call the SA-position $i$ a representative of the edge $\edge(i)$ for all $i < |T|$. For example, $0$, $2$, $7$, and $10$ are all representatives of the edge $\{0, 1, 2, 3\}$ in Figure~\ref{figure:hypercsa:construction:psi_jumps}. 

For $v \in V$, let $S(v) = \{\select_D(v+1, 1), \dots, \select_D(v+2, 1)-1\}$ be the set of the SA-positions $i$ where $T[SA[i]] = v$. For example, for $v=1$, we have $S(1) = \{2, 3, 4\}$. By construction of the (compressed) suffix array, every $S(v)$ is a contiguous set of natural numbers, as can be seen in Figure~\ref{figure:hypercsa:construction:psi_jumps}.

HyperCSA uses the $\delta$-code by Elias~\cite{elias1975} on the differences $\Psi[i] - \Psi[i-1]$ between two values to encode $\Psi$. This works well because for node $v \in V$, $\Psi[S(v)]$ is a contiguous increasing integer sequence~\cite{adler2026}. However, for this paper, we will work with the numerical $\Psi$ without any encoding.

\section[State of the Art]{State of the Art on Updateable Hypergraph Compression}

Many hypergraph compression approaches do not have an update algorithm for their compressed structure. This includes the approaches HybHuff of Zhao et al.~\cite{zhao2025} and the Hygra implementation using Ligra by Shun~\cite{shun2020}. Liu et al.~\cite{liu2024} propose an approach to reorder the elements of an incidence list and thereafter compress it. In comparison to the other mentioned methods, Liu et al. provides a maintenance algorithm that can be used to update the order of elements in the incidence lists after modification. The worst-case runtime of their compression algorithm and their maintenance algorithm is \[\bigO\left(\sum_{u \in V} (mdeg(u))^2 + n^2\right)\] with $mdeg(u)$ being the sum of the ranks of hyperedges that are incident to the node $u \in V$. Regardless of the type of modification, their approach always uses the same maintenance algorithm. In Table~\ref{table:hypercsa:modify:theoretical_runtime_comparison}, we show all worst-case runtime bounds for the operations of HyperCSA and compare them to the worst-case runtime bounds for the incidence list. We highlight that every update operation of HyperCSA has a lower worst-case bound than the approach of Liu et al.~\cite{liu2024}.

\section[Dynamic HyperCSA]{Dynamic HyperCSA}\label{section:hypercsa}

\sidenode{Overview}
We present the modification operations in the following order:
\begin{enumerate}
    \item deleting an edge,
    \item removing a node from an edge,
    \item inserting a node to an edge,
    \item adding an edge,
    \item deleting a node, and
    \item adding a node.
\end{enumerate}
We remark that modifying the HyperCSA structure normally renders all prior by queries obtained representatives useless.

\begin{table}
    \centering
    \begin{tabular}{p{\dimexpr 0.245\linewidth-2\tabcolsep}p{\dimexpr 0.36\linewidth-2\tabcolsep}p{\dimexpr 0.395\linewidth-2\tabcolsep}}
        method & incidence list & HyperCSA \\
        \hline
        construct & $\bigO(M_G)$ & $\bigO(|E_G|R_G(\lg(R_G)+\lg(|E_G|)))$ \\

        $\edge(p)$ & $\bigO(\rank(\edge(p)))$ & $\bigO(\rank(\edge(p)))$ \\

        $\decompress(G)$ & $\bigO(M_G)$ & $\bigO(M_G)$ \\\hline

        $\degree(v)$ & $\bigO(1)$ & $\bigO(1)$ \\

        $\rank(i)$ & $\bigO(1)$ & $\bigO(\rank(i))$ \\

        $\contains(v)$ & $\bigO(\degree(v))$ & $\bigO(\degree(v))$ \\

        $\contains(Q)$ & $\bigO(n + R_G\min_{i}(\degree(v_i)) )$ & $\bigO(n + R_G\min_{i}(\degree(v_i)))$ \\

        $\argexists(Q)$ & $\bigO(n\min_{i}(\degree(v_i)))$ & $\bigO\left(\sum_{i=0}^{n}\log(\degree(v_i))\right)$ \\\hline

        $\deleteEdge(i)$ & $\bigO(R_G \cdot \Delta_G)$ & $\bigO(M_G)$ \\
        $\removeNode(n, i)$ & $\bigO(\rank(i) + \degree(n))$ & $\bigO(M_G + R_G^2 \cdot \Delta_G)$ \\

        $\insertNode(n, i)$ & $\bigO(\rank(i) + \degree(n))$ & $\bigO(M_G + R_G^2 \cdot \Delta_G)$ \\

        $\addEdge(e)$ & $\bigO(\rank(e))$ & $\bigO(M_G)$ \\

        $\deleteNode(n)$ & $\bigO(M_G)$ & $\bigO(\degree(n) \cdot (M_G + R_G^2 \cdot \Delta_G))$ \\

        $\addNode(n)$ & $\bigO(1)$ & $\bigO(1)$ \\

    \end{tabular}
    \caption[Comparison of runtimes of the incidence list and HyperCSA.]{Comparison of worst-case runtimes of the incidence list and HyperCSA. Runtimes of queries are included from Adler et al.~\cite{adler2026}. The space requirements are $2 M_G + |E_G| + |V_G|$ integers for the incidence list and $M_G (H_0(T) + 1)+ \bigO(M_G \log(H_0(T)))$ bits for HyperCSA (with $H_0(T)$ being the zero-order empirical entropy of $T$). We use $Q$ as a shortcut for $\{v_1, \dots, v_n\}$.}\label{table:hypercsa:modify:theoretical_runtime_comparison}
\end{table}

We will use our example from Figure~\ref{figure:example_hypergraph} of a hypergraph $H$ containing the hyperedges $\{2\}$, $\{2\}$, $\{0, 1, 2, 3\}$, $\{1, 2, 3\}$, and $\{0, 1, 2, 4\}$. We have $D  = 1 0 1 0 0 1 0 0 0 0 1 0 1 1$ and $\Psi = [2, 4, 7, 8, 9, 5, 6, 10, 11, 12, 0, 3, 1]$.

\subsection{Deleting an Edge}\label{section:hypercsa:modify:remove_edge}

\sidenode{Position versus Representative}
In a list of hyperedges, deleting the third edge is easy and unambiguous. However, in the HyperCSA data structure, it is not clear, which edge is the third. Therefore, any representative $p$ of an edge $\edge(p)$ can be used to call $\deleteEdge(p)$.

We delete an edge containing position $p$ in two steps: First, we mark all positions we are going to delete. Second, we shorten the HyperCSA data structure. Both steps can be done for multiple deletions at once, but we present the algorithm for only one deletion. 

\sidenode{$Z$}
First, we create a bitvector $Z$ with $Z[i] = 1$ if $i = \Psi^k[p]$ for some $k$ and $Z[0] = 0$ otherwise. Thereby, $Z$ encodes the positions of the edge $\edge(p)$, which are the positions we delete from HyperCSA. 

As an example, we show $\deleteEdge(3)$ on the hypergraph $H$ of Figure~\ref{figure:hypercsa:construction:psi_jumps}. Because $\Psi^1[3] = 8$, $\Psi^2[3] = \Psi[8] = 11$, and $\Psi^3[3] = \Psi[11] = 3$, we get $Z = 0 0 0 1 0 0 0 0 1 0 0 1 0$. In the following paragraphs, we use $D'$, $\Psi'$, and $\Psi''$ for presenting and unambiguously referencing steps of our approach. However, our implementation does not generate new arrays for them. Our implementation alters the existing $D$ and $\Psi$ array.

\sidenode{Update $D$}
The second step is to compute a new $D'$ from $D$: The length of $D'$ is the length of $D$ minus $\rank(e)$, the rank of the edge we want to delete, which is the length of the cycle containing $p$. Retrieving $D'$ is easy: We always append the next value to $D'$, so we build $D'$ left to right starting at position $0$. We iterate over $Z[i]$ and $D[i]$: If $Z[i] = 0$, we append $D[i]$ to $D'$. If $Z[i] = 1$, we skip the value of $D[i]$ except if $D[i] = 1$, because $D[i] = 1$ indicates the start of the interval of the next node $v$. In this case, we add a $1$ and skip the following position if it fulfills $D[i+1] = 0$. If $D[i+1] = 1$, our call $\deleteEdge(p)$ requires reducing the degree of the node $v$ to zero. Deleting the $1$ of its span would cause all larger node IDs to shrink by $1$, which would alter the data significantly. There is no way to compensate for this in HyperCSA, because this deletion violates the minimality of $V$. Therefore, we abort the deletion with an exception in this case. We discuss a solution in Section~\ref{section:hypercsa:modify:unconnected_nodes}.

In our example, we get $D' = 1 0 1 0 1 0 0 0 1 1 1$. In words, the nodes $0$ and $1$ occur two times, the node $2$ is connected to four hyperedges, and the node 3 and 4 to only one hyperedge each. The information about which node is connected to which hyperedge is saved using $\Psi$.

\sidenode{Update $\Psi$}
The third step is to obtain a new $\Psi'$. This works similar to computing $D'$ from $D$, but is even more straightforward: We iterate over $Z[i]$ and $\Psi[i]$ at the same time and we append $\Psi[i] - \rank_Z(\Psi[i], 1)$ to $\Psi'$ if and only if $Z[i] = 0$. Adjusting $\Psi[i]$ by subtracting $\rank_Z(\Psi[i], 1)$ is necessary because we deleted $\rank_Z(\Psi[i], 1)$ positions before position $\Psi[i]$ and we need to compensate this in $\Psi'$.

For example in Step $i = 2$, we have $\Psi[2] = 7$. 
\begin{align*}
    \Psi'[2] &= \Psi[2] - \rank_Z(\Psi[2], 1)\\
    &= 7 - \rank_Z(7, 1) \\
    &= 7 - 1 = 6.
\end{align*}
Overall, we obtain $\Psi' = [2, 3, 6, 7, 4, 5, 8, 9, 0, 1]$.

\sidenode{Runtime Analysis}
Clearly, all data structures use $\bigO(M_G)$ space and we only iterate once over the deleted edge, $D$, and $\Psi$, so deleting the edge takes $\bigO(M_G)$ time.

\subsection{Removing a Node from an Edge}

\sidenode{Determine Position}
The call $\removeNode(p, v)$ supposes us to delete the node $v$ from the edge $e = \edge(p)$. As an example, we will show $\removeNode(3, 2)$. First, we check if $e$ contains $v$. This is achieved by iterating over the positions of the edge starting at $p$ by applying $\Psi$ recursively on the result. For such a position $i$, we determine the corresponding node $\node(i)$ and check if $\node(i) = v$. If we reach $p$ again and we have not found the node $v$, there is no node to delete and we can finish. If we found $v$, we also acquired position $q$ representing the connection of edge $e$ to node $v$. In our example, we retrieve the edge at positions $(3, 8, 11)$, and $q = 8$.

\sidenode{Update $D$}
Next, we calculate the new $D'$ from $D$. Mostly, we copy the values from $D$, but we skip the value at position $q$ if $D[q]=0$. If $D[q]=1$ and $D[q+1]=0$, we put a $1$ at $D'[q]$ and skip the position $q+1$ in $D$. If $D[q] = D[q+1] = 1$, deleting $q$ reduces the degree of $v$ to zero. Like for deleting edges, this is not allowed and we abort. We discuss a solution in Section~\ref{section:hypercsa:modify:unconnected_nodes}.

In the example, $\removeNode(3, 2)$ means we remove node $2$ from the edge $(1, 2, 3)$, which results in the new edge $(1, 3)$. We adjust $D$ to $D' = 1 0 1 0 0 1 0 0 0 1 0 1 1$, where we deleted one $0$ after the third $1$, which means that the node $2$ occurs one fewer time.

\sidenode{Update $\Psi$}
The third step is adjusting $\Psi$. The idea is simple: The jump pointing to $q$ should now point to $\Psi[q]$, so the cycle skips position $q$ afterwards. The main work on adjusting $\Psi$ results from the small shifts of positions. We generate $\Psi'[i]$ by iterating over $\Psi[j]$, where the indices might differ because we skip position $q$, so $j = i$ for $i < q$ and $j = i + 1$ for $i \geq q$.
\begin{enumerate}
    \item If $\Psi[j] < q$, we set $\Psi'[i] = \Psi[j]$.
    \item If $\Psi[j] > q$, we set $\Psi'[i] = \Psi[j] - 1$. 
\end{enumerate}
Both cases are just a simple copy of $\Psi$ to $\Psi'$ considering that we skip position $q$. If $\Psi$ points to a position after $q$, we subtract one, because all such positions will be shifted once to the left.
\begin{enumerate}[resume]
    \item If $\Psi[j] = q$ and $\Psi[\Psi[j]] < q$, we set $\Psi'[i] = \Psi[\Psi[j]]$
    \item If $\Psi[j] = q$ and $\Psi[\Psi[j]] > q$, we set $\Psi'[i] = \Psi[\Psi[j]] - 1$
\end{enumerate}
These two cases handle the position $j$ with $\Psi[j] = q$, which is the position we want to skip. Obviously, only one of the two cases occurs per computation of $\Psi'$, and this case applies to only one position, so they are the last two cases tested due to the efficiency. We skip the value $q$, which is why we use $\Psi[\Psi[j]]$. In principle, the case distinction is the same as in the previous two cases: If the next position is higher, we need to subtract one for the shift. The latter two cases also answer the question, whether we delete the highest node of the edge, because we always get the case that $\Psi[\Psi[j]] > q$ unless $v$ is the highest node.

\sidenode{Incorrectly Sorted Interval}
In the example, we get \[\Psi' = [2, 4, 7, 10, 8, 5, 6, 9, 11, 0, 3, 1]\] from \[\Psi = [2, 4, 7, 8, 9, 5, 6, 10, 11, 12, 0, 3, 1].\] At position $q' = 4$ where $\Psi[q'] = q = 8$, we retrieve for $\Psi'[4]$ the value $\Psi[\Psi[4]] = 11$ with the fourth case from above. Because $11 \geq 8$, we decrease $11$ by $1$, so we get $\Psi'[4] = 10$. Also, all references to positions greater than $8$ are reduced by $1$. But this has the negative effect that the subsequence $7, 10, 8$, which belongs to node $1$, is no longer a contiguous increasing subsequence. Therefore, we now explain the function for modifying $\Psi'$ such that the returned $\Psi''$ fulfills the contiguous increasing subsequences property again. The following operations will be used by other operations like $\insertNode(\cdot)$ as well. Also, these steps are the last steps necessary for completing $\removeNode(\cdot)$.

\subsection{Interval Upward and Downward Sort}

\sidenode{Setting}
Assume that we are given a new $D'$, $\Psi'$, a position $p$ and the information that $p$ is the only position that violates the contiguous increasing subsequence property. In this section, we will explain the algorithm for the case that $\Psi'[p] > \Psi'[p+1]$, which we call \textit{interval upward sort}. The case $\Psi'[p] > \Psi'[p-1]$ works analogously and is called \textit{interval downward sort}. From the modifications to the HyperCSA structure, we always know if we need the upward or downward sort algorithm. We know by $\Psi'[p] > \Psi'[p+1]$ that the value $\Psi'[p]$ should have a higher position within the interval $S(v)$ of the node $v$ that belongs to position $p$. This is the reason we call this modification the \textit{interval upward} sort.

In our example, we get $D' = 1 0 1 0 0 1 0 0 0 1 0 1 1$, $\Psi' = [2, 4, 7, 10, 8, 5, 6, 9, 11, 0, 3, 1]$, and $p = 3$. We also know that the subsequence $[7, 10, 8]$ is not sorted correctly and that the $10 = \Psi[3]$ needs to be moved to a higher position within the interval $S(1)$. 

\sidenode{Find Correct Position}
First, we find the position $q$ to switch the value of $\Psi'[p]$ to, which is the position within $S(v)$ such that $\Psi'[q] < \Psi'[p] < \Psi'[q+1]$. We find the position $q$ by scanning $S(v)$ starting at $p$ upwards towards the end of $S(v)$. Note, that if the scanning results in $p = q$, then everything is already sorted correctly. Also, finding $q$ can be done in at most $|S(v)| = \degree(v)$ steps. In our example, we get $q = 4$.

\sidenode{Resort}
Second, we move every value in $(p, q]$ to the next smaller position and move the value $\Psi'[p]$ to $q$. We call the newly updated function $\Psi''$. This sorts the interval $S(v)$ correctly. This step obviously also needs only $\degree(v)$ steps. However, it creates the problem that all positions $z$ with $\Psi'[z] \in [p, q] \subseteq S(v)$ need to be updated, because we changed the order of these values. We fix this with the steps three and four.

In our example, $(p, q] = (3, 4]$, so we only swap the values at the positions $3$ and $4$. Thus, we get $\Psi' = [2, 4, 7, 8, 10, 5, 6, 9, 11, 0, 3, 1]$. This joins the two cycles of the edges $\edge(3)$ and $\edge(4)$, so we need to adjust the jumps to the swapped values.

\sidenode{Correct Shifted Values}
Third, we fix the jumps $\Psi'[z]$ towards the values in $(p, q]$. Note that we shifted the intended destination of the jumps down by one position, so the values $\Psi'[z]$ need to be decreased by one. We find such a position $z$ with $\Psi'[z] \in (p, q]$ by iterating over $\Psi''$ until $(\Psi'')^k[y] \in (p, q]$ for a $y \in [p, q)$. We know $k = \rank(e)$ for the edge $e$ that belongs to position $y$, because the cycle of the edge is correct except for the last jump at $z = (\Psi'')^{k-1}[y]$. This means, we get the position $z$ in at most $R_G$ steps. We then set $\Psi''[z] = \Psi'[z]-1$.
This small decrease in value cannot cause a new violation of the contiguous increasing subsequence property at the positions $z$, because all values in $\Psi'$ are unique and, by a decrease by one and restoring the uniqueness in step four by only one adjustment, the increasing order of the values around $z$ cannot have changed, as this would require a change of at least two. A consequence of the order at positions $z$ with $\Psi''[z] \in [p, q)$ not causing a new violation is, that we can update the values and do not have any branches of additional necessary updates other than step four. Thereby, we only have at most $\degree(v)$ edges to update and need at most $\bigO(R_G)$ steps for such an update.

In our example, we only need to fix the jump to position $4$, which is now at position $3$. We therefore traverse from the value starting at $\Psi''[3]$ until we get to position $z$ with $\Psi''[z] = 4$, which requires the lookups at $8$, $11$ and $1$. Then, we set $\Psi''[1] = 4-1 = 3$, so we have $\Psi'' = [2, 3, 7, 8, 10, 5, 6, 9, 11, 0, 3, 1]$. Note that we have a second $3$ here, which we will update in the next step.

\sidenode{Loop}
Fourth, for the value $z_p$ with $\Psi'[z_p] = p$, we set $\Psi''[z_p] = q$. Like in step three, we use the partial cycle in $\Psi''$ containing $q$ to get to position $z_p$ and then set $\Psi''[z_p] = q$. Note that we increased the value from $p$ to $q$ here. This means that $z_p$ may now be a position that violates the contiguous increasing subsequence property in the same way $p$ did before our adjustment. Thus, we set $p = z_p$ and restart with the first step. 

In our example, we get $z_p = 10$ with $\Psi''[z_p] = p = 3$ after just one step by traversing $\Psi''[q] = \Psi''[4] = 10$. Now, we set $\Psi''[z_p] = q = 4$ and get \[\Psi'' = [2, 3, 7, 8, 10, 5, 6, 9, 11, 0, 4, 1].\] We would continue with $p = z_p$. We can end iterating here, because $[0, 4]$ is a contiguous increasing subsequence and we are finished with the interval upward sort.

It is not trivial to see how long we must sort intervals. First, we prove that we need at most $R_G\cdot\Delta_G$ update steps. Then, we will prove that this algorithm needs at most $2R_G$ such update steps in case of removing or inserting a node to an edge.

\begin{theorem}
    \sidenode{General Case}
    Interval upward sort terminates after at most $R_G\cdot\Delta_G$ iterations. Its total runtime is in $\bigO(R_G^2 \cdot \Delta_G^2)$.
\end{theorem}

\begin{proof}
    First, if $e$ is the edge that belongs to position $p$ of the first iteration, all subsequent values of $p$ are also representatives of edge $e$. 
    The first $p$ can thereby only move to the end of the interval $S(v)$ and, in the worst-case, is only moved one step per visit of the interval $S(v)$. After at most $R_G$ steps, we visit $S(v)$. Thus, after at most $R_G \Delta_G$ iterations, we aborted early or we have moved $p$ to the end of $S(v)$.

    The runtime of each iteration is dominated by the third step, which takes $\bigO(R_G \cdot \Delta_G)$ steps. As we have at most $R_G \cdot \Delta_G$ iterations, the worst-case time is $\bigO(R_G^2 \cdot \Delta_G^2)$.\qed
\end{proof}

\begin{theorem}
    \sidenode{Special Case of Adding or Removing}
    We get an upper bound of $2R_G$ iterations if the interval upward or downward sort is called by removing or adding a single node. In this case, the interval upwards or downwards sort's worst-case runtime is $\bigO(R_G^2 \cdot \Delta_G)$.
\end{theorem}

\sidenode{Special Case of Insertion or Deletion of Node from Edge}
We can get an upper bound of $2R_G$ iterations if the interval upward or downward sort is called by removing or adding a single node. Then, interval upwards or downwards sort's worst-case runtime is $\bigO(R_G^2 \cdot \Delta_G)$.

\begin{proof}
    First, if $e$ is the edge that belongs to position $p$ of the first iteration, all subsequent values of $p$ also represent the same edge $e$. Next, we consider the suffixes of $T$: The move from position $p$ to $q$ sorts $e$ towards the edges that have at least the same first symbol of the suffix, due to the value of $\Psi$. So, after $\rank(e)$ steps, we update $p$ inductively correctly by all $\rank(e)$ symbols of its suffix. The former suffix during the construction is no longer necessary to consider as we used it only to obtain the contiguous increasing subsequence property in the first place. Thus, the reference of $p$ is moved to the correct position at step $\rank(e) + 1 \leq R + 1$. The second round of updates is adjusting the numbers according to this now corrected position. The next visit of $S(v)$ does not require any further changes in the interval.\qed
\end{proof}

The overall result of $\removeNode(3, 2)$ is $\Psi'' = [2, 3, 7, 8, 10, 5, 6, 9, 11, 0, 4, 1]$ and $D' = 1 0 1 0 0 1 0 0 0 1 0 1 1$.

\subsection{Inserting a Node to an Edge}\label{section:hypercsa:modify:add_node}

\sidenode{Inverse Operation}
With the call $\insertNode(p, v)$, we add the node $v$ to the edge $\edge(p)$. As an example, we will show $\insertNode(4, 2)$ on the result of the $\removeNode(3, 2)$ example, so we have $D = 1 0 1 0 0 1 0 0 0 1 0 1 1$ and $\Psi = [2, 3, 7, 8, 10, 5, 6, 9, 11, 0, 4, 1]$ initially. The call $\insertNode(4, 2)$ will revert the change done by $\removeNode(3, 2)$, so the result will be $D' = 1 0 1 0 0 1 0 0 0 0 1 0 1 1$ and $\Psi'' = [2, 4, 7, 8, 9, 5, 6, 10, 11, 12, 0, 3, 1]$. We highlight that the representative $3$ for the removal and the representative $4$ for the insertion represent the same edge, but the actual value was changed due to the modifications made during the removal in the last section.

\sidenode{Determine Surounding Positions}
First, we need to find the pair of positions $(i, j)$ such that $\Psi(i) = j$ and $\node_D(i) < v < \node_D(j)$. If such a pair does not exist, because for all nodes $v' \in \edge(p)$, we have either $v' < v$ or $v' > v$, we want the pair $(i, j)$ such that $\Psi(i) = j$ and $\node_D(i) > \node_D(j)$. During this traversal, if we already have $v \in \edge(p)$, we abort the insertion early. 

\sidenode{Determine Insert Position}
The next major goal is to insert $j$ into the subsequence $\Psi[S(v)]$. Therefore, we first search the position $q \in S(v) \cup \{\max(S(v))+1\}$ with $\Psi[q-1] < j < \Psi[q]$ (to simplify, we assume $\Psi[-1]=-1$ and $\Psi[M_G]=M_G$ for the boundaries here). Because $\Psi[S(v)]$ is a contiguous increasing subsequence, we can use a binary search to find $q$. 

In the example, we have $(i, j) = (4, 10)$. We search the position $q$ in the subsequence $[5, 6, 9, 11]$ where we should put the value $10$ using a binary search. We want to insert the $10$ between $9$ and $11$, so $q = 8$ in $\Psi$.

\sidenode{Update $D$}
Second, we copy $D$ to $D'$, but at the $v$-th span of $D$, we increase the length of that span by one. In the example, we get $D' = 1 0 1 0 0 1 0 0 0 0 1 0 1 1$.

\sidenode{Update $\Psi$}
Third, we do a similar step for $\Psi'$: If both $z, \Psi[z] < q$, we set $\Psi'[z] = \Psi[z]$. If $z < q$ and $\Psi[z] \geq q$, we set $\Psi'[z] = \Psi[z]+1$. If $z \geq q$ and $\Psi[z] < q$, we set $\Psi'[z+1] = \Psi[z]$. If both $z, \Psi[z] \geq q$, we set $\Psi'[z+1] = \Psi[z]+1$. In other words, we move values higher than $q$ up by one to create the space we need for the new node in the edge. We can now set $\Psi'[i] = q$ and $\Psi'[q] = j+1$ and thereby inserted the node $v$ to the edge.

In the example, the first copy steps results in $[2, 3, 7, 9, 11, 5, 6, 10, _, 12, 0, 4, 1]$ and the two explicit assignments produce $\Psi' = [2, 3, 7, 9, 8, 5, 6, 10, 11, 12, 0, 4, 1]$. Now, the subsequence $[7, 9, 8]$ of node $1$ is not ordered properly.

\sidenode{Incorrectly Sorted Interval}
By the binary search, we know that the subsequence of the interval $S(v)$ is properly ordered. In the same way as for deleting a node from an edge, the increase by one cannot violate the contiguous increasing subsequence property, so the only position that might be incorrect is position $i$. Again, we can use the interval downward or upward sort here. We need the interval downward sort if we reduced $\Psi[i]$, as in our example, or the interval upward sort, if we increased $\Psi[i]$. Increasing $\Psi[i]$ happens if and only if we add a new highest node to the edge. In the example, we get $\Psi'' = [2, 4, 7, 8, 9, 5, 6, 10, 11, 12, 0, 3, 1]$ as the result after the interval downward sort.

\sidenode{Runtime Analysis}
The runtime of $\insertNode(p, v)$ is dominated by the steps of copying $D$ to $D'$ and $\Psi$ to $\Psi'$, which both take $\bigO(M_G)$ time, or by the interval upward or downward sort using $\bigO(R_G^2 \cdot \Delta_G)$ time. So, in total we get $\bigO(M_G + R_G^2 \cdot \Delta_G)$ in the worst-case. 

\subsection{Inserting a new Edge}\label{section:hypercsa:modify:add_edge}

\sidenode{Setting}
The call $\addEdge(e)$ inserts a new edge $e$ into the existing hypergraph $G$. Compared to the previous three modifications of the structure, we do not have a representative $p$ to work with. We assume that the nodes in $e$ are sorted, so $e[0] < e[1] < \dots < e[\rank(e)-1]$. As an example, we will insert the edge $\{1, 2, 3\}$ into the hypergraph with the edges $\{2\}$, $\{2\}$, $\{0, 1, 2, 3\}$, and $\{0, 1, 2, 4\}$, which is given by $\Psi = [2, 3, 6, 7, 4, 5, 8, 9, 0, 1]$ and $D = 1 0 1 0 1 0 0 0 1 1 1$. This will undo the step of $\deleteEdge(3)$ in Section~\ref{section:hypercsa:modify:remove_edge}.

\sidenode{Update $D$}
For $e$, we need one new position for every node $e[i]$. More precisely, for $i < \rank(e)$, we insert one new element in every subsequence in $\Psi[S(e[i])]$. Using this information about the location of the newly inserted elements, we compute $D'$ from $D$ as follows: Let $D'$ be a bitvector of length of $M_G+1 + \rank(e)$. We iterate over $D$ and set the current position of $D'$ to the bit occurring at $D[i]$ and we skip one position of $D'$ if $D[i] = 1$ and $\node_D(i) \in e$. Because the nodes in $e$ are ordered, we will find the $e[i]$ in order of their index $i$. This results in $D' = 1 0 1 0 0 1 0 0 0 0 1 0 1 1$ from $D = 1 0 1 0 1 0 0 0 1 1 1$, where we increased the length of the spans associated with the nodes $1$, $2$, and $3$.

\sidenode{Determine Insert Position}
The HyperCSA structure was constructed using suffix sorting. To get the insert positions for the nodes of $e$, we will use this construction idea. The problem is that we modified $\Psi_{CSA}$ to $\Psi$: From any position $p$ in $\Psi$, we only know a prefix of the suffixes of the construction string $T$. 
For example, for any edge $e'$, the longest exactly known prefix of a suffix starts at $e'[0]$. On the contrary, we know the construction rules of $T$, so we know that after edge $e'$ was an edge $e''$ with $e' > e''$ or $e'$ was the lexicographically smallest edge.

\sidenode{Compare to Parts of Edges}
Our goal is to find the insert position $p_0$ of $e[0]$ in $S(e[0])$. We will use a binary search on the interval $S(e[0])$ and compare $e$ to the current position. This position might not be a first node of an edge: In our example, the interval $S(e[0]) = S(1) = [2, 3]$ contains representatives to the edges $\{0, 1, 2, 3\}$, and $\{0, 1, 2, 4\}$, which both do not have the node $1$ as their first node. Thus, we always compare the edge $e$ to a part $s$ of another edge $e'$ that contains the node $e[0]$.

\sidenode{Edge-Suffix}
Let $e'$ be another edge that contains $e[0]$. We only consider the part $s$ of $e'$ that consists exactly of the nodes of $e'$ larger or equal to node $e[0]$. Therefore, we call $s$ an \emph{edge-suffix}. We refer with $\rank(s)$ to the length of $s$. We end up in one of several cases:

\begin{itemize}
    \item If \sidenode{Cases for Suffix Comparison}$e[m] < s[m]$ for an $m < \min(\rank(e), \rank(s))$ and $e[i] = s[i]$ for all $i < m$: $e$ has a smaller suffix than $s$.
    \item If $e[m] > s[m]$ for an $m < \min(\rank(e), \rank(s))$ and $e[i] = s[i]$ for all $i < m$: $e$ has a larger suffix than $s$.
    \item If $e[i] = s[i]$ for all $i < \min(\rank(e), \rank(s))$:
    \begin{itemize}
        \item If $\rank(e) < \rank(s)$: $e$ has a smaller suffix than $s$. 
        \item If $\rank(e) > \rank(s)$: $e$ has a higher suffix than $s$.
        \item If $\rank(e) = \rank(s)$:
        \begin{itemize}
            \item $e \neq e'$: $e$ has a higher suffix than $e'$.
            \item $e = e'$: we can choose, because both decisions will give us the same result.
        \end{itemize}
    \end{itemize}
\end{itemize}  

The first two cases are the classic suffix sorting cases, so the first unequal node determines the order of the suffixes. Next, if all nodes are equal, one suffix can be longer than the other, which is the case if the ranks of $e$ and $s$ are different. Let us consider the case $\rank(s) > \rank(e)$ and $e[i] = s[i]$ for all $i < \rank(e)$: The next node in $s$ is larger than the previous and the next node $n$ after the last node of $e$ is smaller or equal to its previous node due to the construction of $T$, so $s[\rank(e)] > s[\rank(e)-1] = e[\rank(e)-1] \geq n$. Thus, $e$ has the smaller suffix. The case $\rank(s) < \rank(e)$ works analogously. Here, the special case is if $\rank(e) = \rank(s)$, so we have $e = s$. If $s \neq e'$, there is a node $n \in e'$ that is strictly smaller than any node in $e$ and we get $e > e'$. In $T$, the suffix after $e$ is higher than the suffix after $e'$ by the downward sort of the edges and thereby we have that $e$ has the higher suffix. If $e' = s$, then $e' = e$. This means, their cycles would be next to each other in $\Psi$. In this rare case, we can use the cycle of $e'$ to get all positions of $e'$ easily, either all are before or all after the cycle of $e'$. This results in the same $\Psi$ in both cases.

In the example of inserting $e = \{1, 2, 3\}$, the interval is $S(e[0]) = S(1) = [2, 3]$ and we first compare $e$ to the edge containing position $3$. Because $\Psi[3] = 7 \in S(2)$, we also have $e[1] = s[1]$. Next, $\Psi[7] = 9 \in S(4)$, so we have $e[2] = 3 < 4 = s[2]$, so $e$ has the smaller suffix. By the binary search, we now compare to $\Psi[2] = 6$. As $6 \in S(2)$ and $\Psi[6] = 8 \in S(3)$, we did $\rank(e)$ comparisons. Because $\Psi[8] = 0 < 8$, we also have reached the end of the edge-suffix $s$ here, so $\rank(e) = \rank(s)$. If $e' = s$, then we should have $\Psi[8] \in S(1)$, but we have $\Psi[8] = 0 < S(1)$, so $e > s$. Thereby, the insert position is $3$, which means we insert $e$ before the edge $e$ at position $p_0 = 3$.

\sidenode{Runtime Analysis}
Using the S-type position to determine the end of the edge-suffix $s$ and the fact that the implication $e' \neq s \Rightarrow e' \neq e$ requires $e = s$, we can efficiently obtain the position of $e[0]$ without further extension of the cases above. To conclude, we can decide within at most $R_G$ steps if $e$ or $s$ has the higher suffix and we need at most $\bigO(\log(\Delta_G))$ steps for the binary sort, so we can get the insert position $p_0$ of $e[0]$ in $\bigO(R_G \cdot \log(\Delta_G))$ steps. 

\sidenode{Determine Position of Remaining Nodes}
For the other nodes, we use the search from $\insertNode(\cdot)$: We know the position $p_0$ of $e[0]$, so we can search in the subsequence $\Psi[S(e[\rank(e)-1])]$ for position $q$ with the smallest value higher than or equal to $p_0$ using a binary search. We set $p_{\rank(e)-1} = q$, because we want the contiguous increasing subsequence property. $\Psi[S(e[\rank(e)-1])]$ fulfills the property, and we will set $\Psi'[p_{\rank(e)-1}] = p_0$. Having $p_{\rank(e)-1}$, we can search for $p_{\rank(e)-2}$. So we iterate downward to search the needed positions until we find $p_1$. For the purpose of storing and using the information best, we use a bitvector $Z$ of length $|D|=M_G+1$ that is initially $0$, and we set $Z[p_i] = 1$ for the found insert positions $p_i$ for all $i < \rank(e)$.

In our example, we saw that $p_0 = 3$. As $\rank(e) = 3$, we next search the position $p_2$. Thus, we search in $\Psi$ in the interval $S(e[2]) = S(3) = [8, 9)$ for the value $p_0 = 3$ using a binary search. Because $\psi[8] = 0$, we get $p_2 = 9$. Thereafter, we continue for $p_1$: Here, $S(2) = [4, 8)$ and we get $p_1 = 7$, because $\Psi[6] = 8$, $\Psi[7] = 9$, and we searched for the value $p_2 = 9$. To conclude, we do not get the values $p_i$, but we get $Z = 0 0 0 1 0 0 0 1 0 1 0$.

\sidenode{Expand $\Psi$}
Given $Z$, the construction of $\Psi'$ is straightforward: We iterate over $\Psi$ with index $i$, and if at the current position $i$ the value $Z[i]$ is $1$, we skip a position in $\Psi'$, so the index $j$ of $\Psi'$ is increased by an additional step. This produces the yet unassigned positions in $\Psi'$ for the new edge $e$. Additionally, we need to adjust the value $\Psi[i]$ before we can set it to $\Psi'[j]$. We must increase $\Psi[i]$ by the number of added unassigned positions up to $\Psi[i]$, so we set $\Psi'[j] = \Psi[i] + \rank_Z(\Psi[i]+1, 1)$, where the $1$s in $Z$ are exactly the number of unassigned positions. We increment $\Psi[i]$ by one, because if $Z[i] = 1$, we also need to consider that $Z[i] = 1$ means shifting this character by an additional position for a new unassigned position. We remark that we use a data structure to answer the rank queries in $\bigO(1)$ time, which highlights the advantage of encoding all insert positions $p_i$ by a single bitvector $Z$.

In the example, we use the bitvector $Z = 0 0 0 1 0 0 0 1 0 1 0$ to transform \[\Psi = [2, 3, 6, 7, 4, 5, 8, 9, 0, 1]\] into \[\Psi = [2, 4, 7, \_, 9, 5, 6, 10, \_, 12, 0, \_, 1].\] The underscores stand for the unassigned insert positions.

\sidenode{Update $\Psi$}
As the last step of inserting edge $e$, we assign values to the unassigned positions of the new edge. We use $p_{i-1} = \select_Z(i, 1) + i$ for $1 \leq i \leq \rank(e)$ to get the position of node $e[i-1]$. So, we set $\Psi'[p_{i-1}] = i_{i}$ for all $0 \leq i < \rank(e)$ and $\Psi'[p_{\rank(e-1)}] = p_0$.

In the example, we get $\Psi' = [2, 4, 7, 8, 9, 5, 6, 10, 11, 12, 0, 3, 1]$. Together with $D' = 1 0 1 0 0 1 0 0 0 0 1 0 1 1$, we correctly inserted $e = \{1, 2, 3\}$ and reverted the changes.

\sidenode{Inverse Operation}
Overall, for any edge $e \subseteq V$, we have \[(\Psi', D') = \addEdge(\Psi, D, e)\] and \[\deleteEdge(\Psi', D', \argexists(\Psi', D', e)[0]) = (\Psi, D),\] so we can revert any addition of a new edge. Also, if $e = \edge(p)$, we have for any representative $p$, \[\Psi', D' = \deleteEdge(\Psi, D, p),\] and \[(\Psi, D) = \addEdge(\Psi', D', e),\] so the deletion is invertible if and only if we are allowed to delete the edge that belongs to $p$. We extend this by a special call $\argexists(\Psi, D, e)$, which gives us a set of edge representatives of edges that are equal to the edge represented by $p$. We remark that we are not allowed to delete an edge if we thereby delete the last connection of a node. In the next paragraph, we present an idea to overcome this issue. 

\subsection{Unconnected Nodes in HyperCSA}\label{section:hypercsa:modify:unconnected_nodes}

\sidenode{Motivation}
With $\deleteEdge(\cdot)$ or $\removeNode(\cdot)$, we can reach the case that we want to delete the last connection of a node $n$. In this chapter, we present an easy solution to that problem. \sidenode{Bitvector $N$}Let $N$ be a bitvector of length $|V_G|$. If the node $n \in V$ has at least one incident hyperedge $e$, then $N[n] = 1$. Up to this point, this was a requirement, but we allow nodes $n$ that are not connected to any hyperedge by setting $N[n] = 0$. Consequently, we can obtain the node degrees as
\[\degree(n) = \begin{cases}
    0 & N[n] = 0 \\
    |S(\rank_{N}(n, 1))| & N[n] = 1
\end{cases}.\]

\sidenode{Delete a Node}
Thus, we can now delete the last connection of a node $n$ by setting $N[n] = 0$. This occurs while deleting a span from $D$ that had only length $1$, so if we had $D[i] = D[i+1] = 1$ for $i = \select_D(n, 1)$. 

\sidenode{Adjustment to Queries}
Allowing unconnected nodes needs a small adjustment for queries and for modifications of HyperCSA. First, we look at the queries: Given a sequence $s$ of nodes $n_i$ for any query, we check for every node $n_i$ in $s$ that $N[n_i] = 1$, otherwise the query obviously has no matching edge by definition of $N$. If all nodes are connected, we need to translate them, because there might be lower nodes that are not connected to any edge. We set $n_i' = \rank_N(n_i, 1)$ and $s'$ to be the sequence of the $n_i'$. The $\rank(\cdot)$ reduces $n_i$ by the number of zeros in $N$ before $n_i$. We can then perform the query using $s'$. Without the translation, we search within the wrong spans of $D$.

\sidenode{Adding a Node}
We adjust the values of nodes using the transformation from $n_i$ to $n_i'$ of the previous paragraph. In the case of adding at least one node $n$ to the structure that has $N[n] = 0$, we set $N[n] = 1$ and add a new span of length $1$ to $D$ at the right location besides the changes to $\Psi$ described in Section~\ref{section:hypercsa:modify:add_node} or Section~\ref{section:hypercsa:modify:add_edge}.

\sidenode{Adding a new Node}
It is now also possible to add a new node to $G$ by appending a zero to $N$. If the node should be inserted in between other nodes, we shift the values of higher positions in $N$ to the next higher position. We delete a node $n$ from $G$ that has $N[n] = 0$, by deleting the $n$-th bit in $N$, so we shift the bits with a position higher than $n$ to the position one smaller and shorten the length of $N$ by one. If we want to delete a node with $N[n] = 1$, we first detach all connected edges from that node. A naive solution calls $\removeNode(\select_D(n, 1), n)$ for $S(n)$ times. Each such call will detach one edge from the node $n$. This method is naive, because we could detach all edges from the node $n$ at the same time and reduce the update cost for the remaining parts of $D$ and $\Psi$, which dominates the runtime. We do not present an advanced algorithm here, because in our opinion, deleting a connected node happens rarely in real-world scenarios.

\section{Experimental Evaluation}

Our tests were performed on a Debian 5.10.209-2 machine with 512GB RAM and 128 Cores Intel(R) Xeon(R) Platinum 8462Y+ @ 2.80GHz. We evaluated the social network community hypergraphs CoFr and CoOr, which are available from \href{https://snap.stanford.edu/data/index.html}{SNAP}~\cite{standford_datasets_2014}. The statistics of the datasets are given in Table~\ref{table:hypercsa:evaluation:data_sets}. 

\begin{table}
    \centering  
    \begin{tabular}{p{3.5cm}RRRRR}
        hypergraph $G$ & $|V_G|$ & $|E_G|$ & $M_G$ & $R_G$ & $\Delta_G$ \\ \hline

        \href{https://snap.stanford.edu/data/com-Friendster.html}{com-friendster} (CoFr) & 7,944,949 & 1,620,991 & 23,479,217 & 9,299 & 1,700 \\

        \href{https://snap.stanford.edu/data/com-Orkut.html}{com-orkut} (CoOr) & 2,322,299 & 15,301,901 & 107,080,530 & 9,120 & 2,958 \\ 
    \end{tabular}
    \caption[Datasets of HyperCSA's evaluation.]{Used datasets from \href{https://snap.stanford.edu/data/index.html}{SNAP}~\cite{standford_datasets_2014}.
    }\label{table:hypercsa:evaluation:data_sets}
\end{table}

Construction of the CoFr dataset takes $3$ minutes and $33.951$ seconds. Deleting an edge from CoFr in the HyperCSA format takes only about $26.8349$ seconds. We added an edge within about $27.1503$ seconds. Inserting a node takes about $38.0888$ seconds. Attempting to delete a node from an edge to which the node is not connected takes only $7.366$ milliseconds.

The construction of the CoOr dataset takes about $51$ minutes. Deleting an edge takes about $93.056$ seconds. We add a new edge to the HyperCSA-compressed CoOr hypergraph in about $93.2257$ seconds. Attempting to delete a node from an edge to which the node is not connected takes only $17.381$ milliseconds, whereas adding a new node to an edge takes about $126.184$ seconds.

The results of both CoFr and CoOr show that our update-algorithms are significantly faster than a decompress-update-compress approach, which needs the time to compress the datasets and the updates have only a negligible influence on the compression time. If multiple updates are needed, for example deleting multiple edges at a time, these steps can be combined to reduce the overall time needed.

\section[Conclusion]{Conclusion}

In this paper, we introduced algorithms to add and remove nodes or edges and algorithms to add or remove single nodes from edges in a hypergraph that was compressed using HyperCSA. We provided worst-case runtimes for all presented algorithms and showed by a practical evaluation that the algorithms are much faster than the naive decompress-update-compress approach.

\bibliographystyle{psc}
\bibliography{biblio}

\end{document}